\documentclass[reprint,nofootinbib,amsmath,amssymb,aps,floatfix,superscriptaddress,twocolumn]{revtex4-2}
\usepackage{graphicx}
\usepackage{dcolumn}
\usepackage[colorlinks,linkcolor=magenta,anchorcolor=cyan,citecolor=blue]{hyperref}
\usepackage{bm}
\usepackage[multiple]{footmisc}

\begin{document}
\title{Magnetic Reconnection Heating of Thin Accretion Disks around Kerr Black Holes}

\author{Zhen Li} 
\email{zhen.li@just.edu.cn}
\affiliation{School of Science, Jiangsu University of Science and Technology, Zhenjiang 212100, China}

\date{\today}

\begin{abstract}
The magnetic reconnection process near a black hole can efficiently convert magnetic energy into the kinetic and thermal energy of plasma outflows, enabling the extraction of rotational energy from a spinning black hole. However, its impact on the surrounding accretion disk remains insufficiently explored. In this work, we develop a framework to investigate the 'heating' effect of magnetic reconnection outflow plasma on thin accretion disks around Kerr black holes. We incorporate the energy and angular momentum fluxes of reconnection outflow into the disk conservation laws as additional source terms, yielding a modified radiative flux. We investigate the dependence of the modified radiative flux on the black hole spin, reconnection radius, plasma magnetization, and outflow orientation. For a fixed magnetic field prescription, the contribution of the reconnection outflow plasma to the modified radiative flux decreases as the black hole spin increases. Its dependence on the reconnection radius differs inside and outside the ergosphere: inside, a smaller radius enhances disk heating through more efficient extraction of black hole rotational energy, whereas outside, a larger radius produces stronger heating due to reduced radial dilution of the outflow. The reconnection outflow plasma contribution also increases with plasma magnetization and the radial component of the outflow velocity. Our results demonstrate that magnetic reconnection can provide an additional source of energy and angular momentum for relativistic accretion disks, establishing a connection between plasma processes near rotating black holes and the observable accretion disk properties, providing a new perspective on the strong gravitational field regime.
\end{abstract}

\maketitle

\section{Introduction}\label{sec1}

Black holes are among the most fundamental predictions of general relativity and provide a unique arena where strong gravity and high energy astrophysical processes exist simultaneously. Recent observations of black hole images also provide important opportunities to probe the underlying spacetime and accretion physics \cite{shadow1, shadow2}. Therefore, understanding the physical processes occurring in the vicinity of black holes is essential for establishing a connection between relativistic gravitational theory and astrophysical observations.

A particularly important component of the black hole environment is the accretion disk. When matter falls toward a black hole, the released gravitational energy can be converted into thermal and radiative energy, giving rise to a luminous accretion disk. The thin accretion disk model provides one of the most widely used frameworks for describing the radiation produced by matter accreting to a black hole \cite{disk1,disk2,disk3}. In this framework, the radial transport of mass, angular momentum, and energy determines the local dissipation rate and consequently the radial distribution of the radiative flux. For a sufficiently optically thick and geometrically thin disk, the emitted radiation carries direct information about the dynamics of the accreting matter and the properties of the central spacetime. The physics of accretion disks is therefore closely connected to a broad range of black hole observables \cite{ts0, ts1, ts3, ts4, ts5, ts6, ts15, ts16, ts21, ts23}. Moreover, the emission from the accretion disk provides the photon source to the formation of black hole images \cite{t1s7, t1s8, t1s9, t1s10, t1s11, t1s12, t1s13, t1s14, t1s17, t1s18, t1s19, t1s20, t1s22, t1s24, t1s25, t1s26}. Consequently, a reliable interpretation of black hole observations requires not only an appropriate spacetime model but also a physically consistent description of the processes governing the accretion disk.

In addition to the accretion disk, magnetic fields play a crucial role in determining the dynamics of plasma in the vicinity of a black hole. In particular, magnetic reconnection provides an efficient mechanism for converting magnetic energy into plasma kinetic and thermal energy, thereby contributing to a variety of high energy astrophysical phenomena, such as relativistic jets \cite{bz}. The dynamics of magnetic reconnection around a rotating black hole are particularly interesting because the process can provide a channel for extracting the rotational energy of the black hole \cite{MR,MR0}. 

Rotating black holes can give rise to an antiparallel magnetic field configuration in the equatorial plane \cite{mr1,mr2}, and such configurations have been found in numerical simulations of magnetized accretion flows as well as in studies of black hole imaging observations \cite{mr3,mr4,mr5,mr6,shadow1,shadow2}. When the current sheet formed between oppositely directed magnetic field lines becomes sufficiently elongated, its aspect ratio can exceed a critical value, triggering fast magnetic reconnection \cite{mr7,mr8,mr9}. Numerical simulations further also indicate that magnetic reconnection is typically localized around a point in the current sheet \cite{mr3,mr4,mr5,mr6}, commonly referred to as the X-point \cite{MR}, where magnetic field lines reconnect and the plasma is efficiently accelerated. This process can occur both inside and outside the ergosphere of a rotating black hole. Magnetic energy stored in the reconnecting field is then converted into the kinetic and thermal energy of the plasma. Magnetic reconnection generally produces two oppositely directed plasma outflows, with one component being accelerated and the other decelerated. Inside of ergosphere, if the decelerated plasma acquires negative energy with respect to an observer at infinity, while the accelerated outflow carries an energy flux exceeding the contributions associated with its rest mass and thermal energy, the rotational energy of the black hole can be extracted through the reconnection process \cite{MR}. The frame dragging effect of a rapidly rotating black hole can sustain recurrent magnetic reconnection events.

Although the role of magnetic reconnection in accelerating relativistic plasma and extracting black hole rotational energy has been extensively investigated \cite{mrr1, mrr2, mrr3, mrr4, mrr5, mrr6, mrr7, mrr8, mrr9, mrr10, mrr11, mrr12, mrr13, mrr14, mrr15, mrr16, mrr17, mrr18, mrr19, mrr20, mrr21, mrr22, mrr23, mrr24, mrr25, mrr26, mrr27}, its subsequent interaction with the accretion disk has received comparatively less attention. In particular, the energy and momentum carried by the reconnection outflows do not necessarily disappear after leaving the reconnection layer. As the outflows propagate away from the X-point, they can transport a finite energy and angular momentum flux through the surrounding spacetime and eventually interact with the accretion disk. A fraction of this transported energy can then be deposited into the disk, providing an additional source of energy beyond the gravitational dissipation described by the standard thin accretion disk model \cite{ns1, ns2}. For convenience and to provide an intuitive description of this process, we refer to it as Magnetic Reconnection Heating (MRH) of the thin accretion disk. Such an interaction naturally raises an important question: how does the energy and angular-momentum deposition associated with magnetic reconnection modify the radial structure of the radiative flux of a relativistic thin accretion disk? A self-consistent treatment of this problem requires the energy-momentum flux of the reconnection outflow to be connected explicitly to the conservation laws governing the thin accretion disk.

In this work, we develop a framework for describing the MRH of a thin accretion disk around a Kerr black hole. We begin by constructing the energy-momentum tensor of the reconnection outflow. Then the interaction between the reconnection outflow and the disk is subsequently incorporated into the total energy-momentum conservation law through corresponding energy and angular-momentum source terms. By integrating the modified conservation equations, we obtain the generalized conservation laws for rest mass, angular momentum, and energy in the presence of MRH. These equations lead to a modified expression for the radiative flux of the thin accretion disk. This formulation therefore provides a direct physical connection between the spacetime geometry of the Kerr black hole, the dynamics of magnetic reconnection, and the observable radiative properties of the accretion disk.

The remainder of this paper is organized as follows. In Sec.~\ref{sec2}, we construct the energy-momentum tensor of the magnetic reconnection outflow in the Kerr spacetime. In Sec.~\ref{sec3}, we derive the modified conservation laws and radiative flux induced by MRH. In Sec.~\ref{sec4}, we systematically examine the effects of the black-hole spin, reconnection radius, plasma magnetization, and outflow orientation on the resulting radiative flux. Finally, Sec.~\ref{sec5} presents our conclusions on this work.

\section{The energy-momentum tensor of magnetic reconnection process}\label{sec2}

In this section, we formulate the energy-momentum tensor governing the magnetic reconnection process and derive the corresponding energy-momentum tensor of the accelerated outflow plasma. This provides the basis for investigating how the energy and momentum carried by the reconnection outflow are transferred to and affect the surrounding thin accretion disk.

For a rotating black hole, we adopt the Kerr metric to describe its spacetime geometry. In Boyer–Lindquist coordinates $(t,r,\theta,\phi)$, the line element can be written as
\begin{equation}
ds^{2}=g_{tt}dt^{2}+2g_{t\phi}dtd\phi+g_{\phi\phi}d\phi^{2}+g_{rr}dr^{2}+g_{\theta\theta}d\theta^{2},
\end{equation}
where the nonzero metric components are given by
\begin{align}
g_{tt}&=\frac{2Mr}{\Sigma}-1, \qquad g_{t\phi}=-\frac{2aM\sin^{2}\theta}{\Sigma},\nonumber\\
g_{\phi\phi}&=\frac{A}{\Sigma}\sin^{2}\theta, \quad\quad\;\; g_{rr}=\frac{\Sigma}{\Delta}, \quad g_{\theta\theta}=\Sigma,
\end{align}
with the quantities defined as
\begin{align}
\Sigma &= r^{2}+a^{2}\cos^{2}\theta,\\
\Delta &= r^{2}-2Mr+a^{2}, \\
A &= (r^{2}+a^{2})^{2}-a^{2}\Delta\sin^{2}\theta.
\end{align}
Throughout this work, we adopt geometrized units with $G=c=1$.

During the magnetic reconnection process, the following one-fluid energy-momentum tensor can be used to describe the coupled dynamics of the plasma and electromagnetic fields,
\begin{equation}
\overset{\scriptscriptstyle MR}T{}^{\mu\nu} = w U^\mu U^\nu+P g^{\mu\nu} + F^\mu{}_\delta F^{\nu \delta} - \frac{1}{4} g^{\mu\nu}F^{\tau \delta}F_
{\tau\delta},
\end{equation}
where $w$, $P$, and $U^\mu$ denote the enthalpy density, pressure, and four-velocity of the plasma, respectively, and $F^{\mu\nu}$ is the electromagnetic field tensor.

We consider the case of highly efficient magnetic reconnection, in which most of the magnetic energy is converted into the kinetic energy of the plasma, while the electromagnetic energy measured at infinity is negligible compared with the hydrodynamic energy carried by the plasma. Under this assumption, the electromagnetic contribution to the energy-momentum tensor can be neglected. The energy-momentum tensor therefore reduces to
\begin{equation}\label{setbl}
\overset{\scriptscriptstyle MR}T{}^{\mu\nu} \approx w U^\mu U^\nu+Pg^{\mu\nu}.
\end{equation}
The plasma energy density is most conveniently evaluated in the locally non-rotating frame, commonly referred to as the zero-angular-momentum-observer (ZAMO) frame \cite{bard}. In the ZAMO frame, the Kerr metric takes the locally Minkowskian form
\begin{equation}
ds^2 \equiv -d\hat{t}^2 + \sum_{i=1}^3 (d\hat{x}^i)^2 = \eta_{\mu\nu} d\hat{x}^\mu d\hat{x}^\nu,
\end{equation}
where a hat is used to denote quantities defined in the ZAMO frame. The ZAMO coordinates are related to the Boyer-Lindquist coordinates through the following transformations,
\begin{equation}
d\hat{t} = \alpha dt,\qquad d\hat{x}^i = \sqrt{g_{ii}} dx^i - \alpha\beta^i dt,
\end{equation}
where we define
\begin{align}
\alpha &= \left(-g_{tt} + \frac{g_{\phi t}^2}{g_{\phi\phi}}\right)^{1/2} = \left(\frac{\Delta\Sigma}{A}\right)^{1/2} , \\
\beta^\phi &= \frac{\sqrt{g_{\phi\phi}}\,\omega^\phi}{\alpha} = \frac{\omega^\phi}{\alpha}\left(\frac{A}{\Sigma}\right)^{1/2}\sin\theta,
\end{align}
and $\omega^\phi = -g_{\phi t}/g_{\phi\phi} = 2M^2 a r / A$ is the angular velocity of the frame dragging. 

In the ZAMO frame, the energy-momentum tensor of the expelled plasma Eq.(\ref{setbl}) is given by
\begin{equation}\label{zamot}
\overset{\scriptscriptstyle MR}T{}^{\hat\mu \hat\nu} \approx w U^{\hat\mu} U^{\hat\nu}  + P\eta^{\hat\mu\hat\nu},
\end{equation}
where $U^{\hat\mu} = (\hat\gamma, \hat\gamma \hat{v}{} )$ denotes the four-velocity of the expelled plasma measured in the ZAMO frame, and $\hat v{} = (\hat v{}^r,\hat v{}^\theta,\hat v{}^\phi)$ represents its three-velocity, with radial, polar, and azimuthal components. The corresponding Lorentz factor is given by $\hat\gamma = 1/\sqrt{1-\hat v{}^2}$.

The energy-momentum tensor of the expelled plasma in the two coordinate frames can be related through the tetrad transformation
\begin{equation}\label{bztrans}
\overset{\scriptscriptstyle MR}T{}^{\mu\nu} = e^\mu{}_{\hat a }e^\nu{}_{\hat b} \overset{\scriptscriptstyle MR}T{}^{\hat a\hat b} ,  
\end{equation}
where the tetrad components are given by
\begin{equation}
e^\mu _{\hat a } = 
\begin{pmatrix}
\frac{1}{\alpha} & 0 & 0 & 0 \\
0 & \frac{1}{\sqrt{g_{rr}}} & 0 & 0 \\
0 & 0 &  \frac{1}{\sqrt{g_{\theta\theta}}} & 0 \\
\frac{\beta^\phi }{\sqrt{g_{\phi\phi}}}& 0 & 0&  \frac{1}{\sqrt{g_{\phi\phi}}}
\end{pmatrix}
\end{equation}
Here, we focus on two important components, $\overset{\scriptscriptstyle MR}T{}^{tr}$ and $\overset{\scriptscriptstyle MR}T{}^{\phi r}$, which will be used to derive the radiative flux in the next section. Their corresponding mixed-variance tensors can be written as
\begin{align}
\overset{\scriptscriptstyle MR}T{}_t{}^r &= g_{tt}\overset{\scriptscriptstyle MR}T{}^{t r} + g_{t\phi} \overset{\scriptscriptstyle MR}T{}^{\phi r}, \\
\overset{\scriptscriptstyle MR}T{}_\phi{}^r &= g_{\phi \phi}\overset{\scriptscriptstyle MR}T{}^{\phi r} + g_{\phi t} \overset{\scriptscriptstyle MR}T{}^{t r}.
\end{align}
Using Eqs. (\ref{zamot}) and (\ref{bztrans}), these components can be expressed in terms of the corresponding quantities in the ZAMO frame as
\begin{align}
\overset{\scriptscriptstyle MR}T{}_t{}^r &= g_{tt} \frac{w\hat\gamma{}^2\hat v{}^r}{\alpha\sqrt{g_{rr}}} +g_{t\phi} \frac{w\hat\gamma{}^2\hat v{}^r(\beta^\phi+\hat v{}^\phi)}{\sqrt{g_{\phi\phi}g_{rr}}},\label{tr}\\
\overset{\scriptscriptstyle MR}T{}_\phi{}^r &= g_{\phi\phi} \frac{w\hat\gamma{}^2\hat v{}^r(\beta^\phi+\hat v{}^\phi)}{\sqrt{g_{\phi\phi}g_{rr}}} +g_{t\phi} \frac{w\hat\gamma{}^2\hat v{}^r}{\alpha\sqrt{g_{rr}}}.\label{pr}
\end{align}

Prior to magnetic reconnection, we consider a single, coherent plasma flow in the reconnection region, which we refer to as the bulk plasma. The bulk plasma here denotes the entire pre-reconnection plasma configuration and should not be confused with either of the individual plasma outflows generated by the reconnection process. Once reconnection takes place, this initially coherent plasma flow is effectively divided into two oppositely directed outflows, corresponding to the two sides of the reconnecting magnetic field. The energy and momentum carried by these outflows provide the basis for describing the subsequent interaction between the reconnection products and the surrounding accretion flow.
Assuming that the bulk plasma mediating the reconnection process initially follows a circular orbit around the Kerr black hole in the equatorial plane, its Keplerian angular velocity in Boyer–Lindquist coordinates is given by
\begin{equation}\label{aw}
\Omega_K = \pm \frac{M^{1/2}}{r^{3/2} \pm a M^{3/2}},
\end{equation}
where the plus and minus signs correspond to co-rotating and counter-rotating orbits, respectively. In the following discussion, we consider only the co-rotating case. Circular timelike orbits extend from spatial infinity inward to the limiting circular photon orbit, with the corresponding photon orbit radius given by
\begin{equation}
r_{\rm ph}
=2M\left[1+\cos\left(\frac{2}{3}\arccos(\mp a)\right)\right],
\label{eq:photonorbit}
\end{equation}
where the plus and minus signs correspond to co-rotating and counter-rotating orbits. This provides a lower bound on the radial location of the X-point. Assuming that the bulk plasma follows a circular timelike orbit, the X-point radial coordinate, denoted as $r_0$, must be located outside the limiting circular photon orbit on the equatorial plane, i.e.,
\begin{equation}
r_0 > r_{\rm ph}.
\end{equation}
This condition ensures that the reconnection region lies within the domain where stable or unstable timelike circular motion can be physically realized, thereby allowing the magnetic reconnection process to operate on the orbiting plasma. Consequently, the photon-orbit radius sets the innermost physically admissible location of the X-point in present model.

To analyze the localized magnetic reconnection process and explicitly determine the properties of the expelled plasma, we introduce the local rest frame ${x'^{\mu}}$ of the bulk plasma. We choose the local frame ${x'^\mu}$ such that the $x'^1$ axis is aligned with the radial direction $x'^1 = r$, while the $x'^3$ axis points along the azimuthal direction $x'^3=\phi$. In this local rest frame, the three-velocity of the expelled plasma can be written as
\begin{equation}
\vec v_{out} = (v^r_{out},0,v^\phi_{out}),
\end{equation}
where $v^\theta_{out}=0$ indicates that the outflow is confined to the equatorial plane. The magnitude of the outflow three-velocity is determined by the macroscopic magnetic field strength around the black hole, $B_0$, and is given by \cite{MR}
\begin{equation}
v_{out} = \sqrt{\frac{\sigma_0}{1+\sigma_0}},
\end{equation}
where $\sigma_0 =B_0^2/w_0$ denotes the upstream plasma magnetization of the reconnection layer, and $w_0$ is the upstream enthalpy density. The corresponding Lorentz factor of the expelled plasma in the local rest frame is then given by $\gamma_{out} = 1/\sqrt{1-v^2_{out}}$.

Since the Keplerian velocity of the bulk plasma Eq.~(\ref{aw}) measured in the ZAMO frame, denoted as $\hat v_K$, is given by
\begin{equation}
\hat{v}_K = \frac{A}{\Delta^{1/2}} \left[ \frac{(M/r)^{1/2} - a(M/r)^2}{r^3 - a^2 M^3} \right] - \beta^\phi.
\end{equation}
The corresponding Lorentz factor is $\hat\gamma_K = 1/\sqrt{1-\hat v_K^2}$. We can then express the velocity of the expelled plasma measured in the ZAMO frame, $\hat v$, in terms of the outflow velocity $v_{out}$ in the local rest frame and the Keplerian velocity $\hat v_K$ of the bulk plasma in the ZAMO frame, by applying the special-relativistic velocity addition formulas. Since the bulk plasma is assumed to move on a circular equatorial orbit, the transformation from the local rest frame of the bulk plasma to the ZAMO frame involves only a Lorentz boost in the azimuthal direction. For the co-rotating plasma outflow, the azimuthal and radial components of $\hat v$ are then given by
\begin{align}
\hat v{}^\phi &= \frac{\hat v_K+ v_{out}\cos{\xi}}{1+ v_Kv_{out}\cos{\xi}},\\
\hat v{}^r &=  \frac{v_{out}\sin{\xi}}{\hat\gamma_K(1+ v_Kv_{out}\cos{\xi})},
\end{align}
where $\xi =\arctan(v^r_{out}/v^\phi_{out})$ denotes the orientation angle of the expelled plasma outflow. The corresponding Lorentz factor $\hat \gamma$ in the ZAMO frame can also be obtained from the Lorentz factor $\gamma_{out}$ in the local rest frame and $\hat\gamma_K$ in the ZAMO frame as
\begin{equation}
\hat\gamma = \hat\gamma_K\gamma_{out}(1+ v_Kv_{out}\cos{\xi}).
\end{equation}
With these relations, all the quantities appearing in Eqs. (\ref{tr}) and (\ref{pr}) can be determined explicitly. We can therefore evaluate the contribution of the magnetic reconnection outflow to the thin accretion disk, as discussed in the following section.

\section{The radiative flux of thin accretion disk with MRH}\label{sec3}

As introduced earlier, we assume that the magnetic reconnection process is localized around a dominant X-point. In realistic accretion flows, magnetic reconnection is likely to occur over an extended radial region outside the limiting photon orbit, rather than being confined to a single radial location. In this work, we characterize the distributed reconnection region by an effective radius $r_0$, which can be interpreted as the energy-weighted location of the dominant reconnection sites.

The expelled plasma carries energy and momentum away from the magnetic reconnection region. After reconnection, the plasma propagates outward freely before interacting with the thin accretion disk. During this propagation, the energy-momentum tensor satisfies the conservation equation
\begin{equation}
\nabla_\mu \overset{\scriptscriptstyle MR}  T_\nu{}^\mu= 0.
\end{equation}
Assuming that the reconnection outflow is stationary, axisymmetric, the conservation equations reduce to
\begin{equation}
\partial_r \left(\sqrt{-g} \overset{\scriptscriptstyle MR}T_t{}^r\right)=
\partial_r \left(\sqrt{-g} \overset{\scriptscriptstyle MR}T_\phi{}^r\right) = 0.
\end{equation}
In the equatorial plane, we have $\sqrt{-g} = \Sigma = r^2$, and therefore
\begin{equation}
r^2 \overset{\scriptscriptstyle MR}T_t{}^r =r^2 \overset{\scriptscriptstyle MR}T_\phi{}^r= \mathrm{const},
\end{equation}
which implies that the energy and momentum carried by the expelled plasma at radius $r$, denoted by $\overset{\scriptscriptstyle MR}T_\nu{}^r(r)$, can be expressed in terms of their values at the X-point as
\begin{align}
\overset{\scriptscriptstyle MR}T_t{}^r(r) &= \overset{\scriptscriptstyle MR}T_t{}^r(r_0)\left(\frac{r_0}{r}\right)^2,\label{rdeptr}\\
\overset{\scriptscriptstyle MR}T_\phi{}^r(r) &= \overset{\scriptscriptstyle MR}T_\phi{}^r(r_0)\left(\frac{r_0}{r}\right)^2.\label{rdeppr}
\end{align}
As the expelled plasma propagates outward to a radius $r$ within the disk, it can interact with the disk material and transfer energy and angular momentum to the accretion disk. This additional energy and momentum deposition heat the disk and consequently modifies its radiative flux. 

To investigate more explicitly how the magnetic reconnection outflow modifies the radiative properties of the accretion disk, we employ the conservation law of the total energy-momentum tensor. Before proceeding, we first introduce the energy-momentum tensor of the thin accretion disk. In and around the equatorial plane, we introduce the coordinates $(t,r,z,\phi)$, where the only difference from the Boyer-Lindquist coordinates is that $z$ is used in place of $\theta$ and $z=r\cos\theta$. Here, $z$ denotes the height above the equatorial plane. In this coordinate system, the energy-momentum tensor of the thin accretion disk can be written as \cite{disk2}
\begin{equation}
\overset{\scriptscriptstyle AD}  T{}^{\mu\nu}= \rho_0(1+\Pi)V^\mu V^\nu+t_{\mu\nu} +V^\mu q^\nu+q^\mu V^\nu,
\end{equation}
where $\rho_o$ and $\Pi$ denote the rest-mass density and specific internal energy, respectively. $V^\mu$ is the four-velocity of the particles in the thin accretion disk, while $t_{\mu\nu}$ represents the stress tensor in the averaged rest frame. We impose the corresponding orthogonality relation $V^\mu t_\mu{}^\nu=0$, while $q^\mu$ is the energy-flow vector.

We can likewise express the energy-momentum tensor of the expelled plasma $\overset{\scriptscriptstyle MR}T{}^{\mu\nu}$ in the $(t,r,z,\phi)$ coordinate system and denote it by $S{}^{\mu\nu}$. Then, in the equatorial plane, we have the relation
\begin{equation}\label{strela}
S{}^{\mu\nu} =  J^\mu_{\ \alpha} J^\nu_{\ \beta} \overset{\scriptscriptstyle MR}T{}^{\alpha\beta},
\end{equation}
where $J^\mu_{\ \alpha} = \text{diag}(1,1,-r,1)$ is the Jacobian matrix. Since the magnetic reconnection outflow is constrained within the equatorial plane, $S{}^{\mu\nu}$ is nonzero only at $z=0$. To facilitate the subsequent integration of the energy-momentum conservation law, we represent the source localized on the equatorial plane as $S{}^{\mu\nu}\delta(z)$. 

Then, the total energy-momentum conservation law for the combined system of the expelled plasma and the thin accretion disk at radius $r$ can then be written as
\begin{equation}\label{cl}
\nabla_\mu T^ {\mu\nu} \equiv \nabla_\mu (\overset{\scriptscriptstyle AD} T{}^{\mu\nu}+\eta(r)S{}^{\mu\nu}\delta(z)) = 0,
\end{equation}
where $\eta(r)$ is an efficiency function introduced to account for the fact that the energy carried by the magnetic reconnection outflow is not necessarily transferred completely to the accretion disk. Only a fraction of the reconnection outflow flux remains available for deposition at radius $r$. We therefore multiply $\eta(r)$ to characterize the efficiency of energy and momentum deposition into the disk. We assume the efficiency function takes a power law form with radius
\begin{equation}
\eta (r) = \left(\frac{r_0}{r}\right)^n,
\end{equation}
where $n$ is the index. A larger $n$ corresponds to a more rapid decrease in the MRH energy deposition with radius. In general, $\eta(r)$ can be determined by the underlying physical processes and treated as a free function without affecting the general MRH formulation.

The integral form of the conservation laws is more convenient for analyzing the overall radiative properties of the accretion disk, since the observable quantities of a thin accretion disk are generally averaged over time and azimuth \cite{disk2}. We therefore integrate the differential conservation equation in Eq.~(\ref{cl}) and use Gauss's theorem to convert it into a three-volume integral. This procedure yields three important conservation laws governing the radiative properties of the thin accretion disk in the presence of MRH: conservation of rest mass, angular momentum, and energy.

Assuming the rest mass carried by the expelled plasma is negligible compared with that of the thin accretion disk. Consequently, the rest-mass conservation equation of the thin accretion disk remains unchanged and takes the form \cite{disk2}
\begin{equation}
\dot M_0 = -2\pi\sqrt{-g'}\Gamma V^r,
\end{equation}
where $\dot M_0 $ denotes the rest-mass accretion rate of the thin accretion disk, and $\sqrt{-g'}$ denotes the metric determinant in the $(t,r,z,\phi)$ coordinate system. In and near the equatorial plane, i.e., within the thin accretion disc, it depends approximately only on the radius $r$ \cite{disk2}. $\Gamma$ is the time-averaged surface density of the thin accretion disk,
\begin{equation}
\Gamma =\int_{-H}^H <\rho_0>dz,
\end{equation}
where $H$ denotes the maximum half-thickness of the disk attained during the time interval $\Delta t$, and $<>$ denotes the average of a function $\Psi(x^\mu)$ over the azimuthal angle $\Delta \phi = 2\pi$ and the time interval $\Delta t$, defined as
\begin{equation}
< \Psi(z, r)> \equiv \frac{1}{2\pi \Delta t}\int_{0}^{\Delta t} \int_{0}^{2\pi} \Psi(t, r, z, \varphi) d\varphi dt .
\end{equation}

However, MRH modifies the standard angular-momentum conservation law of the thin accretion disk. We first convert the angular-momentum component of Eq.~(\ref{cl}) to a more useful integral conservation law by integrating over the 3-volume of the disk between radius $r$ to $r+\Delta r$ and over time $\Delta t$. Then applying the Gauss's theorem, we obtain
\begin{widetext}
\begin{align*}
0 &= \int_\mathcal{V} \nabla_\mu T_\phi{}^\mu\sqrt{-g'}dtdrdzd\phi= \int_\mathcal{\partial V}T_\phi\,^\mu d^3\Sigma_\mu\nonumber\\
& = \;\left\{ \int_{-H}^{H}\int_{t}^{t+\Delta t}\int_{0}^{2\pi} \left(\rho_0(1+\Pi)V_\phi V^r +t_\phi{}^r+V_\phi q^r +q_\phi V^r +\eta  S_\phi{}^r \delta(z)\right)\sqrt{-g'} d\phi dtdz\right\}_{r}^{r+\Delta r}\\
&\quad +\left\{ \int_{r}^{r+\Delta r}\int_{t}^{t+\Delta t}\int_{0}^{2\pi} \left(\rho_0(1+\Pi)V_\phi V^z +t_\phi{}^z+V_\phi q^z +q_\phi V^z +\eta S_\phi{}^z\delta(z)\right)\sqrt{-g'} d\phi dtdr\right\}_{-H}^{H}\\
&\quad +\left\{\text{total angular momentum in 3-volume}\right\}_{t}^{t+\Delta t},
\end{align*}
\end{widetext}
where the braces indicate that the enclosed quantity is evaluated at the corresponding boundaries. Under the standard thin-disk approximation \cite{disk2}, we further assume
\begin{equation}
\Pi\approx 0, \,q^r\approx 0,\,q^\phi\approx 0,
\end{equation}
which corresponds to negligible specific heat and heat transport within the plane of the disk. Under these assumptions, the first boundary term becomes
\begin{widetext}
\begin{align}
&\left\{\int_{-H}^{H} 2\pi\Delta t\left[ <\rho_0>V_\phi V^r + <t_\phi{}^r> +<\eta S_\phi{}^r \delta(z)>\right]\sqrt{-g'}dz\right\}_r^{r+\Delta r}\\
&=\left\{2\pi\Delta t\sqrt{-g'} \left[ \Gamma L^+V^r+W_\phi{}^r+\int_{-H}^H<\eta S_\phi{}^r\delta(z) >dz \right] \right\}_r^{r+\Delta r}\\
& =\Delta t\left\{ -\dot M_0L^++2\pi \sqrt{-g'} W_\phi{}^r +2\pi \sqrt{-g'}\int_{-H}^H<\eta S_\phi{}^r\delta(z) >dz \right\}_r^{r+\Delta r} ,
\end{align}
\end{widetext}
where $L^+ \approx V_\phi$, since, after averaging over radius, azimuth, and height, the baryonic matter moves approximately along circular, equatorial geodesics around the black hole. We further define the averaged torque as
\begin{equation}
W_\mu{}^\nu =\int_{-H}^H <t_\mu{}^\nu> dz.
\end{equation}

Since the thin accretion disk approximation requires $V^z=0$ and $t_\phi{}^z=0$, and the expelled plasma is assumed to have no flux across the equatorial plane, i.e., $S_\phi{}^z =0$, the second boundary term reduces to
\begin{align}
&\left\{ \int_r^{r+\Delta r}2\pi \Delta t V_\phi <q^z>\sqrt{-g'}dr\right\}_{-H}^H\nonumber\\
& = 2\Delta t(2\pi \sqrt{-g'}L^+F)\Delta r,
\end{align}
where 
\begin{equation}
F = <q^z(r,z=H)> = <-q^z(r,z=-H)>,
\end{equation}
is time-averaged radiative flux flowing out of upper and lower face of thin accretion disk.

The third boundary term can be neglected compared with the first one under the following assumption \cite{disk2}: there exists a time interval that is (a) sufficiently short such that the exterior geometry of the black hole undergoes negligible variation during this interval, while (b) sufficiently long such that, at any considered radius $r$, the total inward mass flux crossing $r$ during this interval greatly exceeds the characteristic mass contained within the shell extending from $r$ to $2r$.

Combining the above results, we obtain the angular-momentum conservation equation
\begin{align}\label{amc}
&\left[-\dot M_0L^+-2\pi \sqrt{-g'}\left(W_\phi{}^r+\int_{-H}^H<\eta S_\phi{}^r \delta(z)>dz\right)\right]_{, r} \nonumber\\
&= 4\pi \sqrt{-g'}FL^+.
\end{align}

In the same manner, by integrating the differential form of the energy conservation equation, we obtain
\begin{widetext}
\begin{align*}
0 &= \int_\mathcal{V} \nabla_\mu T_t{}^\mu\sqrt{-g'}dtdrdzd\phi= \int_\mathcal{\partial V}T_t\,^\mu d^3\Sigma_\mu\nonumber\\
&= \;\left\{ \int_{-H}^{H}\int_{t}^{t+\Delta t}\int_{0}^{2\pi} -\left(\rho_0(1+\Pi)V_t V^r +t_t{}^r+V_t q^r +q_t V^r +\eta S_t{}^r\delta(z)\right)\sqrt{-g'} d\phi dtdz\right\}_{r}^{r+\Delta r}\\
&\quad +\left\{ \int_{r}^{r+\Delta r}\int_{t}^{t+\Delta t}\int_{0}^{2\pi} -\left(\rho_0(1+\Pi)V_t V^z +t_t{}^z+V_t q^z +q_t V^z +\eta S_t{}^z\delta(z)\right)\sqrt{-g'} d\phi dtdr\right\}_{-H}^{H}\\
&\quad +\left\{\text{total energy momentum in 3-volume}\right\}_{t}^{t+\Delta t},
\end{align*}
\end{widetext}

For the first boundary term, we adopt the same thin-disk approximations,
\begin{equation}
\Pi\approx 0, ,q^r\approx 0,,q^t\approx 0,
\end{equation}
where $q^t\approx 0$ corresponds to neglecting heat transport in the time direction. Under these assumptions, the first boundary term becomes
\begin{widetext}
\begin{align}
&\left\{\int_{-H}^{H} -2\pi\Delta t\left[ <\rho_0>V_t V^r + <t_t{}^r> +<\eta S_t{}^r \delta(z)>\right]\sqrt{-g'}dz\right\}_r^{r+\Delta r}\\
&=\left\{2\pi\Delta t \sqrt{-g'}\left[ \Gamma E^+V^r+W_t{}^r+\int_{-H}^H<\eta S_t{}^r\delta(z) >dz \right] \right\}_r^{r+\Delta r}\\
& =\Delta t\left\{ -\dot M_0E^+-2\pi \sqrt{-g'} W_t{}^r -2\pi \sqrt{-g'}\int_{-H}^H<\eta S_t{}^r \delta(z)>dz \right\}_r^{r+\Delta r} ,
\end{align}
\end{widetext}
where we have used $E^+ \approx - V_t$, since the disk particles are assumed to move along circular orbits after averaging.

For the second boundary term, since thin accretion disk requiring $V^z=0$, $t_t{}^z=0$, and MRH requiring $ S_t{}^z=0$, we obtain
\begin{align}
&\left\{ \int_r^{r+\Delta r}-2\pi \Delta t V_t<q^z>\sqrt{-g'}dr\right\}_{-H}^H\nonumber\\
& = 2\Delta t(2\pi \sqrt{-g'}E^+F)\Delta r.
\end{align}

The third boundary term can also be neglected compared with the first one under the similar assumption as we did in angular momentum part. In the same way, we have
\begin{align}\label{ec0}
&\left[\dot M_0E^++2\pi \sqrt{-g'}\left(W_t{}^r+\int_{-H}^H<\eta S_t{}^r \delta(z)>dz\right)\right]_{, r} \nonumber\\
&= 4\pi \sqrt{-g'}FE^+ .
\end{align}

Based on the orthogonality relation $V^\mu t_\mu{}^\nu=0$, which implies $V^\mu W_\mu{}^\nu =0$, we can obtain
\begin{equation}
W_r{}^t = -\frac{V^\phi}{V^t}W_\phi{}^r = -\Omega W_\phi^r,
\end{equation}
where $\Omega$ denotes the angular velocity of the averaged circular geodesic orbit for disk particles. We can then rewrite Eq. (\ref{ec0}) as
\begin{align}\label{ec1}
&\left[\dot M_0E^+-2\pi \sqrt{-g'}\left(W_\phi{}^r\Omega+\int_{-H}^H<\eta S_t{}^r \delta(z)>dz\right)\right]_{, r} \nonumber\\
&= 4\pi \sqrt{-g'}FE^+ .
\end{align}

We have thus obtained the modified integral conservation laws for a thin accretion disk in the presence of MRH. These conservation laws provide the basis for determining the corresponding radiative flux of the thin accretion disk. In the following, we proceed to calculate the radiative flux modified by the magnetic reconnection process.

First, we need to compute the terms related to the integrals of $S_\phi{}^r$ and $S_t{}^r$ over the $z$-direction, we have
\begin{align}
\int_{-H}^H<\eta S_\phi{}^r >\delta (z)dz &= <\eta S_\phi{}^r> |_{z=0}\nonumber\\
&=\eta S_\phi{}^r|_{z=0} \nonumber\\
&= \eta\overset{\scriptscriptstyle MR}  T{}_\phi{}^r|_{\theta=\pi/2},\\
\int_{-H}^H<\eta S_t{}^r >\delta (z)dz  &= <\eta S_t{}^r >|_{z=0} \nonumber\\
&= \eta S_t{}^r|_{z=0} \nonumber\\
&= \eta\overset{\scriptscriptstyle MR}  T{}_t{}^r|_{\theta=\pi/2},
\end{align}
where $\overset{\scriptscriptstyle MR}{T}{}_t{}^r$ and $\overset{\scriptscriptstyle MR}{T}{}_\phi{}^r$ are given by Eqs.~(\ref{rdeptr}) and (\ref{rdeppr}), respectively, with the radius $r$ now taking the radial location of the accretion disk. For the second equality in both equations, we have assumed that the magnetic reconnection process is statistically stationary and axisymmetric, which ensures that the azimuthal and time averages do not alter the value of the relevant quantities. The third equality follows from Eq.~(\ref{strela}), reflecting the fact that the two coordinate systems coincide at the equatorial plane.

For simplicity, we introduce the following quantities to simplify the integral conservation laws:
\begin{align}
f & \equiv \frac{4\pi \sqrt{-h}F}{\dot M_0},\label{fd}\\
\mathcal{W} & \equiv \frac{2\pi \sqrt{-h}W_\phi{}^r}{\dot M_0},\\
S_1&\equiv \frac{2\pi \sqrt{-h}}{\dot M_0}\eta\overset{\scriptscriptstyle MR}  T{}_\phi{}^r|_{\theta=\pi/2},\\
S_2&\equiv \frac{2\pi \sqrt{-h}}{\dot M_0}\eta\overset{\scriptscriptstyle MR}  T{}_t{}^r|_{\theta=\pi/2}.
\end{align}
where $h \equiv g_{rr}(g_{tt}g_{\phi\phi}-g_{t\phi}^{2})$ is the metric determinant of three-dimensional subspace $(t,r,\phi)$. Then Eqs.~(\ref{amc}) and~(\ref{ec1}) give us
\begin{align}
&\left(L^+-\mathcal{W}-S_1\right)_{, r} = fL^+ , \label{eq1} \\
&\left(E^+ - \Omega \mathcal{W}+ S_2\right)_{, r} = fE^+.\label{eq2}
\end{align}
where we use the identity $\sqrt{-g'} = \sqrt{-h}$ in and around the equatorial plane. Multiplying Eq.~(\ref{eq1}) by $\Omega$ and subtracting it from Eq.~(\ref{eq2}), and then applying the energy-angular-momentum relation for circular orbits \cite{disk2}, $E^+{}_{,r} = \Omega L^+{}_{,r}$, we obtain
\begin{equation}
\mathcal{W} = \frac{f(E^+-\Omega L^+)-S_{2,r}-\Omega S_{1,r}}{-\Omega_{,r}}.
\end{equation}
Substituting this result back into Eq.~(\ref{eq1}) and integrating the resulting differential equation for $f$, we obtain
\begin{align}
&\frac{(E^+-\Omega L^+)^2f}{-\Omega_{,r}} = \frac{(E^+-\Omega L^+)(S_{2,r}+\Omega S_{1,r})}{-\Omega_{,r}}\nonumber\\
&+\int[(E^+-\Omega L^+)(L^+_{,r} -S_{1,r})-L^+(S_{2,r}+\Omega S_{1,r})]dr \nonumber\\
&+ \text{constant}.
\end{align}
The integration constant is determined by imposing the physical condition that, at the innermost stable circular orbit (ISCO), the accreting material leaves the disk and plunges directly into the black hole. Following the standard thin-disk assumption, we therefore take the lower integration limit to be ISCO radius, which is given by \cite{bard}
\begin{equation}
r_{\rm isco} = M\bigl[3 + Z_2 \mp \bigl((3 - Z_1)(3 + Z_1 + 2Z_2)\bigr)^{1/2}\bigr],
\end{equation}
where
\begin{align}
Z_1 &= 1 + (1 - a^2)^{1/3}\bigl[(1 + a)^{1/3} + (1 - a)^{1/3}\bigr],\\
Z_2 &= \bigl(3a^2 + Z_1^2\bigr)^{1/2}.
\end{align}
We also set the integration constant to zero as in \cite{disk2}. This yields
\begin{equation}
f = \frac{B}{A}\Theta(r-r_{isco})-\frac{\Omega_{,r}}{A^2}\int_{r_{isco}}^r [A(L^+_{,r} -S_{1,r})-BL^+]dr ,  
\end{equation}
where we define
\begin{equation}
A\equiv E^+-\Omega L^+,\qquad
B\equiv S_{2,r}+\Omega S_{1,r},
\end{equation}
and $\Theta(r-r_{isco})$ is the Heaviside step function, which vanishes for $r<r_{isco}$ and equals unity for $r\ge r_{isco}$. The Heaviside step function is introduced to ensure that the energy and angular-momentum transport is considered only within the accretion-disk region, $r\geq r_{\rm isco}$. Under the thin-disk approximation, the disk is assumed to terminate at the ISCO, and therefore the transport processes inside $r_{\rm isco}$ are neglected in our model. Based on the definition Eq.~(\ref{fd}), the radiative flux can then be written as
\begin{align}\label{rflux}
F(r) &= \frac{\dot M_0}{4\pi \sqrt{-h}} \frac{B}{A}\Theta(r-r_{isco}) \nonumber\\
&- \frac{\dot M_0}{4\pi \sqrt{-h}} \frac{\Omega_{,r}}{A^2}\int_{r_{isco}}^r [A(L^+_{,r} -S_{1,r})-BL^+]dr.   
\end{align}
This expression describes the radiative flux of the thin accretion disk after accounting for the effects of MRH. Physically, $S_1$ and $S_2$ represent the normalized angular-momentum and energy fluxes transferred from the magnetic reconnection outflow to the accretion disk, respectively. Therefore, $S_1$ modifies the angular momentum conservation of the disk, while $S_2$ acts as an energy source term. In the absence of MRH, i.e., when $S_1 = S_2 = 0$, the above expression reduces to the standard radiative flux formula for a thin accretion disk.

\section{Parameter dependence of MRH}\label{sec4}

When considering the effects of MRH, the resulting radiative flux Eq.~(\ref{rflux}) of thin accretion disk is fully characterized by the following parameters: the black hole mass and spin, $M$ and $a$, respectively; the X-point radius $r_0$; the outflow orientation angle $\xi$; the plasma magnetization $\sigma_0$; and the enthalpy density $w$. Although an additional efficiency-function parameter $n$ is introduced, varying $n$ mainly changes the radial decay of the MRH contribution and does not alter the overall trend. For simplicity, we adopt $n=1$ as a fiducial choice in this work, corresponding to a moderate inverse-radius decline of the effective deposition efficiency. Such power law radial scalings are commonly used to characterize the radial structure of magnetic fields, turbulent stresses, and energy dissipation in accretion flows \cite{r1}. We therefore regard $n=1$ as a phenomenological benchmark rather than a first principles prediction. In the following, we investigate the effects of these parameters on the radiative flux of the thin accretion disk with MRH. For simplicity, we set the black hole mass to $M=1$ throughout the discussion. Accordingly, all physical quantities and parameters are expressed in units of $M$.

\begin{figure}[htbp]
    \centering
\includegraphics[scale=0.4]{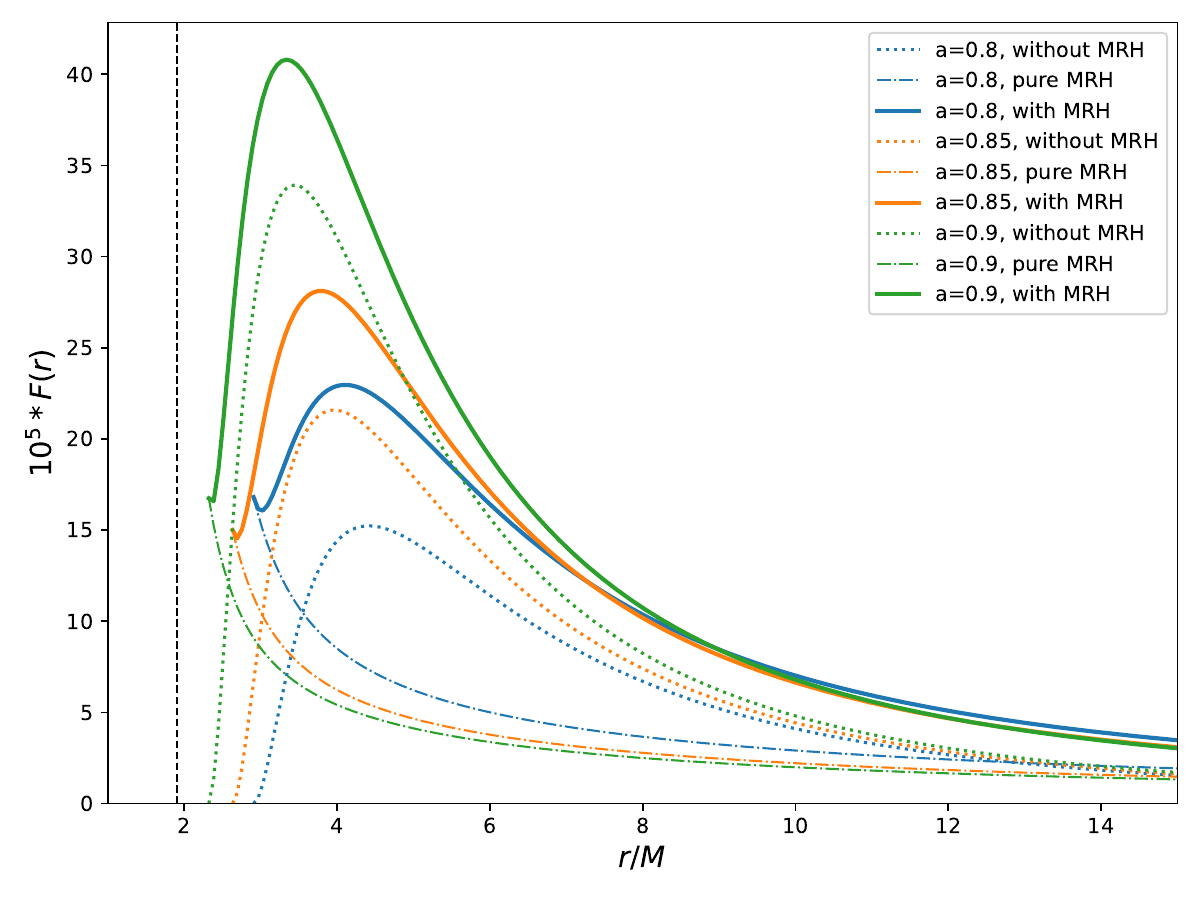}
\caption{The radiative flux $F(r)$ as a function of the radial coordinate $r/M$ for different black hole spins $a/M=0.8,\,0.85,\,0.90$. The solid lines denote the total radiative flux including MRH, the dotted lines denote the radiative flux in the absence of MRH, and the dash-dotted lines represent the contribution due solely to MRH. The X-point location, $r_0/M=1.9$, is marked by the vertical black dashed line. The parameters are fixed at $\xi =\pi/6$, $\sigma_0=100$, and $wM^2=10^{-6}$. All the $F(r)$ values are multiplied by $10^5$ for better illustration.}
\label{a}
\end{figure}

We first investigate the effects of the black hole spin $a$, which characterizes the rotational properties of the Kerr spacetime. In Fig.~\ref{a}, we present the radiative flux $F(r)$ as a function of the radial coordinate $r/M$ for different black hole spins, $a/M=0.8$, $0.85$, and $0.90$. For each spin, we show the total radiative flux with and without MRH, together with the contribution arising solely from MRH. For simplicity, the location of the magnetic reconnection X-point is fixed at $r_0/M=1.9$, as indicated by the vertical black dashed line. The remaining parameters are fixed at $\xi=\pi/6$, $\sigma_0=100$, and $wM^2=10^{-6}$.

As shown in Fig.~\ref{a}, the MRH contribution to the radiative flux decreases as the black hole spin increases, whereas the total radiative flux exhibits an overall increasing trend with $a$. This behavior can be attributed, at least in part, to the enhanced frame-dragging effect in more rapidly rotating Kerr spacetimes. As the spin increases, the stronger frame dragging modifies the kinematics of the reconnection outflow and reduces the fraction of its energy-momentum flux that is efficiently transported in the radial direction toward the accretion disk. Consequently, the energy deposition associated with MRH becomes less efficient for a fixed set of plasma and magnetic-field parameters.

It is important to note, however, that in the present analysis we fix $\sigma_0=100$ and $wM^2=10^{-6}$ for all values of the spin. Since $\sigma_0=B_0^2/w_0$, this prescription effectively keeps the magnetic-field strength $B_0$ fixed when comparing different spins. In realistic accretion systems, however, the magnetic-field strength near the black hole is generally expected to correlate with the black hole spin through the accretion dynamics and magnetic flux accumulation. If such a spin-dependent magnetic-field scaling is taken into account, the increase in $B_0$ with the black hole spin could compensate for, or even overcome, the suppression of the MRH contribution caused by the enhanced frame dragging effect. Therefore, the dependence of the MRH induced radiative flux on the black hole spin may be reversed or substantially modified when a physically motivated spin and magnetic field correlation is incorporated.

\begin{figure}[htbp]
    \centering
\includegraphics[scale=0.4]{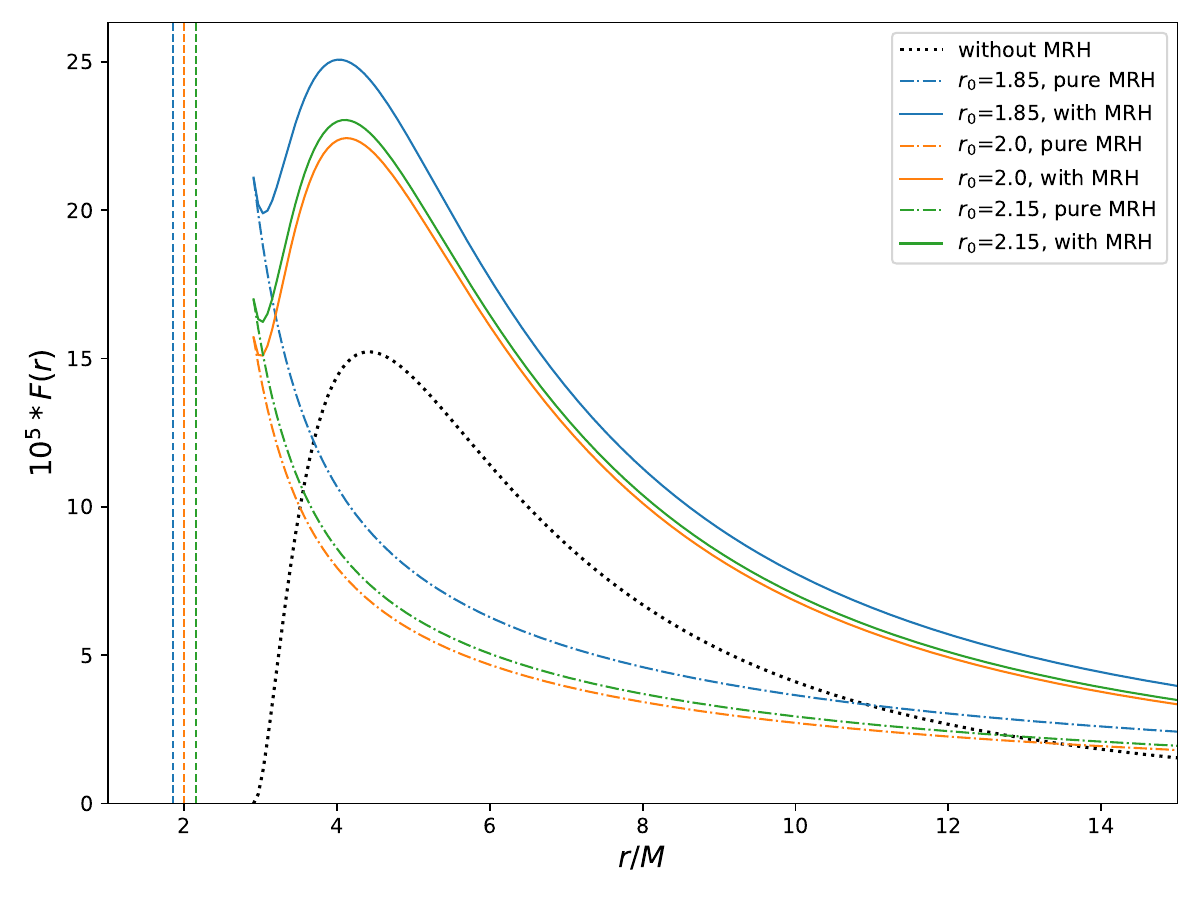}
\caption{The radiative flux $F(r)$ as a function of the radial coordinate $r/M$ for different X-point location $r_0/M=1.85\,,2.00,\,2.15$. The solid lines denote the total radiative flux including MRH, the black dotted line denotes the radiative flux in the absence of MRH, and the dash-dotted lines represent the contribution due solely to MRH. The corresponding X-point locations are indicated by vertical dashed lines with matching colors. The parameters are fixed at $a/M=0.8$, $\xi =\pi/6$, $\sigma_0=100$, and $wM^2=10^{-6}$. All the $F(r)$ values are multiplied by $10^5$ for better illustration.}
\label{r0}
\end{figure}

In Fig.~\ref{r0}, we illustrate the radiative flux $F(r)$ as a function of the radial coordinate $r/M$ for different locations of the magnetic reconnection X-point, namely, $r_0/M=1.85$, $2.00$, and $2.15$. As in the previous analysis, we show the total radiative flux with and without MRH, together with the contribution arising solely from MRH. The corresponding X-point locations are indicated by vertical dashed lines in matching colors. The other parameters are fixed at $a/M=0.8$, $\xi=\pi/6$, $\sigma_0=100$, and $wM^2=10^{-6}$.

As shown in Fig.~\ref{r0}, the dependence of the MRH-induced radiative flux on the X-point location exhibits different behaviors inside and outside the outer boundary of the ergosphere, which is located at $r=2M$ on the equatorial plane. Within the ergosphere, a smaller X-point radius $r_0$ leads to a larger radiative flux. This behavior can be attributed to the more efficient extraction of rotational energy from the black hole when the magnetic reconnection region is located deeper inside the ergosphere \cite{MR}, thereby enhancing the energy flux carried by the reconnection outflow.

Outside the ergosphere, the trend is reversed: a larger X-point radius $r_0$ results in a larger MRH contribution to the radiative flux. This behavior is primarily associated with the radial decay of the energy-momentum flux during its propagation from the reconnection site toward the accretion disk, as described by Eqs.~(\ref{rdeptr}) and (\ref{rdeppr}), together with the radial dependence of the energy-deposition efficiency function. For a fixed observation radius $r$, a larger $r_0$ places the reconnection site closer to the disk region under consideration. Consequently, the energy-momentum flux undergoes less radial dilution and attenuation before reaching the disk, leading to a more efficient deposition of the reconnection energy and, hence, a larger contribution to the radiative flux.

\begin{figure}[htbp]
    \centering
\includegraphics[scale=0.4]{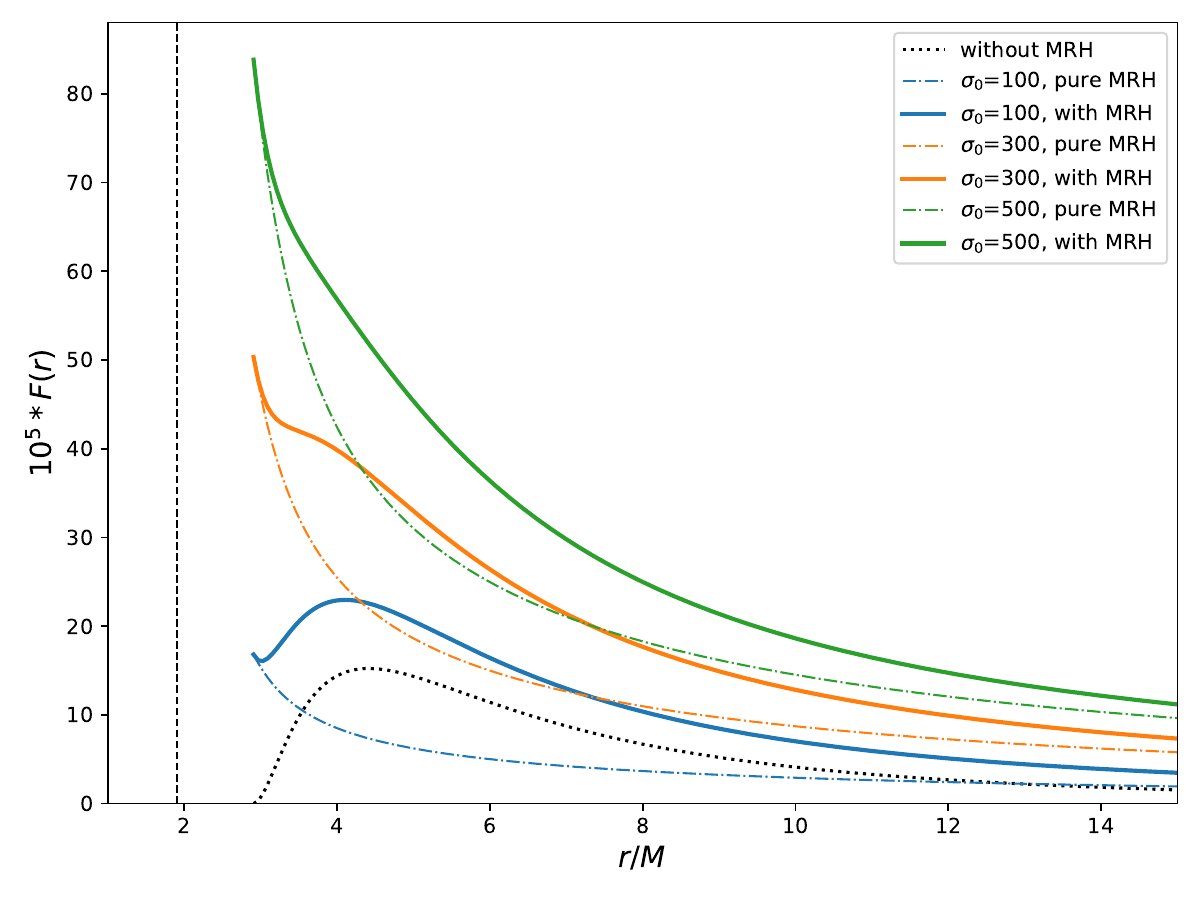}
\caption{The radiative flux $F(r)$ as a function of the radial coordinate $r/M$ for different magnetization parameter $\sigma_0=100,\,300,\,500$. The solid lines denote the total radiative flux including MRH, the black dotted line denotes the radiative flux in the absence of MRH, and the dash-dotted lines represent the contribution due solely to MRH. The X-point location, $r_0/M=1.9$, is marked by the vertical black dashed line. The parameters are fixed at $a/M=0.8$, $\xi =\pi/6$, and $wM^2=10^{-6}$. All the $F(r)$ values are multiplied by $10^5$ for better illustration.}
\label{sigma0}
\end{figure}

We further investigate the dependence of the radiative flux $F(r)$ on the magnetization parameter $\sigma_0$. In Fig.~\ref{sigma0}, we present $F(r)$ as a function of the radial coordinate $r/M$ for $\sigma_0=100$, $300$, and $500$, considering both the cases with and without MRH, as well as the contribution arising solely from MRH. The X-point location is fixed at $r_0/M=1.9$, as indicated by the vertical black dashed line. The remaining parameters are fixed at $a/M=0.8$, $\xi=\pi/6$, and $wM^2=10^{-6}$.

As shown in Fig.~\ref{sigma0}, the radiative flux increases systematically with the magnetization parameter $\sigma_0$. This behavior is physically expected, since $\sigma_0=B_0^2/w_0$,
For a fixed $w_0$, a larger $\sigma_0$ corresponds to a stronger magnetic field, implying a larger amount of magnetic energy available for conversion into the kinetic and thermal energy of the reconnection outflow. Consequently, the enhanced energy flux carried by the outflow leads to a larger energy deposition into the accretion disk and, hence, a stronger MRH contribution to the radiative flux.

It is also worth noting that, in the highly magnetized regime 
$\sigma_0\gg1$, the dependences of Eqs.~(\ref{tr}) and ~(\ref{pr}) on $\sigma_0$ and $w$ become approximately degenerate. Because in this limit, the Lorentz factor of the reconnection outflow satisfies
$\gamma_{\rm out}^{2}=1+\sigma_0\simeq\sigma_0$, while the outflow 
velocity approaches the speed of light, $v_{\rm out}\rightarrow1$. 
Consequently, the dominant factor appearing in Eqs.~(\ref{tr}) and ~(\ref{pr})  scales as
$w\hat{\gamma}^{2}\hat v^r\propto w\sigma_0$.
Therefore, increasing either the magnetization parameter $\sigma_0$ or 
the plasma enthalpy density $w$ enhances the energy and angular momentum 
deposition into the accretion disk in a qualitatively similar manner. Under the highly magnetized prescription adopted in this work, the effect 
of $w$ is thus largely degenerate with that of $\sigma_0$, and we do not 
consider the dependence on $w$ separately for simplicity.

\begin{figure}[htbp]
    \centering
\includegraphics[scale=0.4]{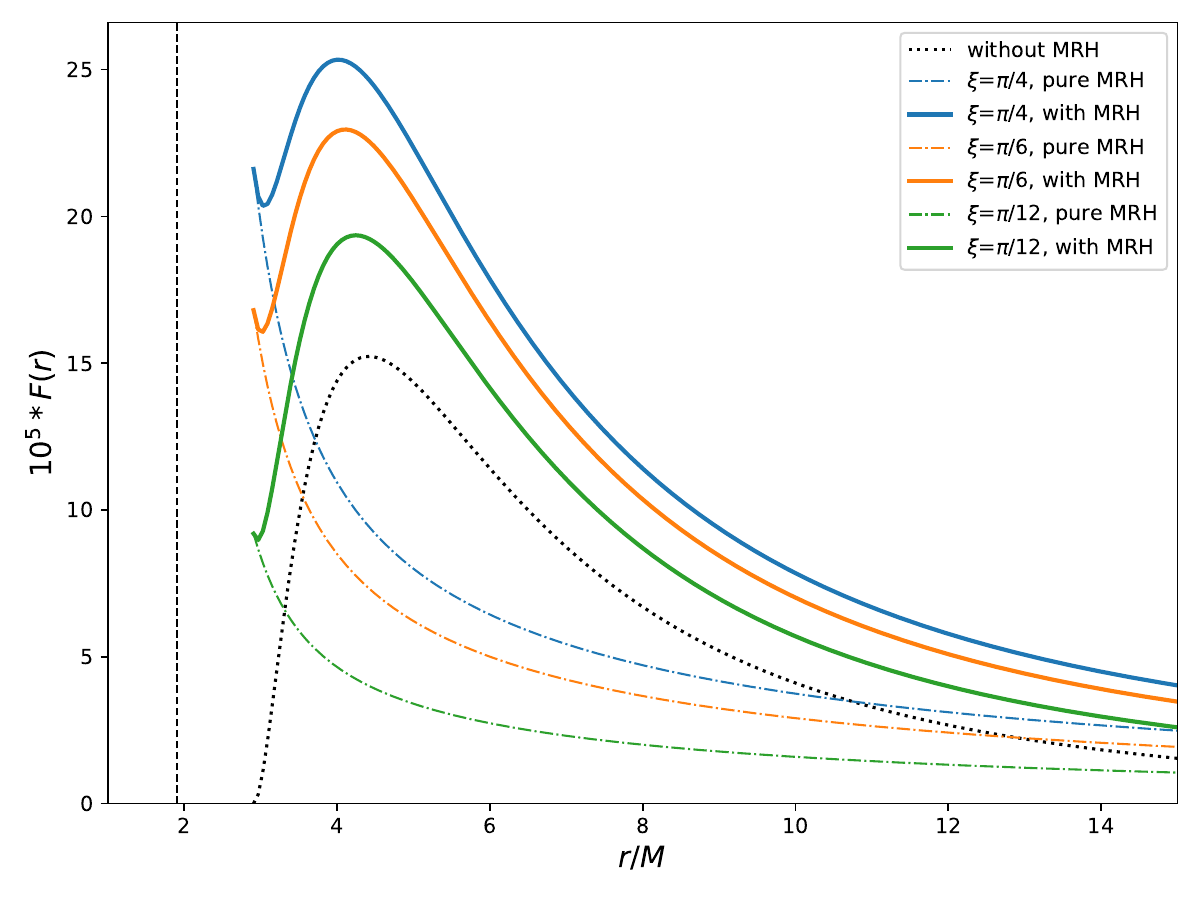}
\caption{The radiative flux $F(r)$ as a function of the radial coordinate $r/M$ for different orientation angle $\xi=\pi/4,\,\pi/6,\,\pi/12$. The solid lines denote the total radiative flux including MRH, the black dotted line denotes the radiative flux in the absence of MRH, and the dash-dotted lines represent the contribution due solely to MRH. The X-point location, $r_0/M=1.9$, is marked by the vertical black dashed line. The parameters are fixed at $a/M=0.8$, $\sigma_0=100$ and $wM^2=10^{-6}$. All the $F(r)$ values are multiplied by $10^5$ for better illustration.}
\label{xi}
\end{figure}

Finally, we investigate the dependence of the radiative flux on the orientation angle $\xi$. In Fig.~\ref{xi}, we present the radiative flux $F(r)$ as a function of the radial coordinate $r/M$ for different orientation angles, $\xi=\pi/4$, $\pi/6$, and $\pi/12$. For each case, we show the total radiative flux with and without MRH, together with the contribution arising solely from MRH. The X-point location is fixed at $r_0/M=1.9$, as indicated by the vertical black dashed line. The other parameters are fixed at $a/M=0.8$, $\sigma_0=100$, and $wM^2=10^{-6}$.

As shown in Fig.~\ref{xi}, the MRH contribution to the radiative flux increases with the orientation angle $\xi$. This behavior can be understood from the geometry of the reconnection outflow. A larger $\xi$ corresponds to a greater radial component of the outflow velocity, allowing a larger fraction of the energy-momentum flux carried by the reconnection plasma to be transported radially toward the accretion disk. As a result, the energy deposition efficiency of the reconnection outflow is enhanced, leading to a larger MRH contribution to the radiative flux.

This behavior is qualitatively consistent with the spin dependence discussed above. While increasing the black hole spin strengthens the frame-dragging effect and tends to modify the direction of the outflow, thereby reducing the efficiency of radial energy transport. Increasing the orientation angle $\xi$ enhances the radial component of the outflow and can partially compensate for the spin effect. Therefore, for a fixed black hole spin, a larger orientation angle generally favors more efficient radial transport of the reconnection energy and consequently produces a stronger MRH contribution to the disk radiative flux.

\section{Conclusion}\label{sec5}

In this work, we have developed a relativistic framework to investigate the heating of a thin accretion disk by magnetic reconnection outflows around a rotating Kerr black hole. We first model the reconnection outflow as relativistic plasma launched from an effective X-point at $r_0$ in the equatorial plane. Its energy-momentum flux is determined by the upstream magnetization $\sigma_0=B_0^2/w$ and the outflow orientation angle $\xi$, with the effects of the orbital motion of the bulk plasma and Kerr frame dragging consistently included through the transformation to the ZAMO frame. Under stationary and axisymmetric conditions, the energy and angular-momentum fluxes carried by the outflow decrease as $r^{-2}$ during propagation. We introduce an effective deposition efficiency $\eta(r)$ and incorporate the resulting energy and angular-momentum fluxes as source terms in the conservation laws of the thin accretion disk. This leads to a modified radiative flux that reduces to the standard relativistic thin accretion disk result when the MRH contribution vanishes.

We find that the MRH-modified radiative flux depends sensitively on the black hole spin, the reconnection radius, the plasma magnetization, and the outflow orientation. For the fixed magnetic field prescription adopted here, the MRH contribution decreases with increasing black hole spin, although this dependence may be substantially modified if the magnetic field strength itself correlates with spin. The dependence on the reconnection radius exhibits a distinct behavior inside and outside the ergosphere. Inside the ergosphere, a smaller $r_0$ produces stronger disk heating because magnetic reconnection can extract black-hole rotational energy more efficiently through frame dragging. Outside the ergosphere, in contrast, a larger $r_0$ enhances the MRH contribution at a fixed disk radius because the outflow undergoes less radial dilution before reaching the disk. The MRH contribution also increases with the plasma magnetization $\sigma_0$ and the orientation angle $\xi$. A larger magnetization provides more magnetic energy for conversion into the kinetic and thermal energy of the reconnection outflow, while a larger $\xi$ enhances the radial component of the outflow velocity and therefore improves the transport of energy and angular momentum toward the disk. These results demonstrate that magnetic reconnection can act as an additional source of energy and angular momentum for relativistic accretion disks, with its impact determined by both the intrinsic energy of the reconnection outflow and its relativistic transport through the Kerr spacetime.

The present framework provides a bridge between magnetic reconnection near rotating black holes and the radiative properties of their accretion disks. Combining the resulting MRH-modified disk flux with relativistic ray-tracing calculations will further allow its consequences for observable quantities, such as thermal spectra, luminosities, and black hole images, to be investigated.

\section*{Acknowledgments}
This work is supported by the National Natural Science Foundation of China (Grant No.~12505073). Zhen Li acknowledges the financial support from the Start-up Funds for Doctoral Talents at Jiangsu University of Science and Technology.
\\
\\


\begin{thebibliography}{100}
\bibitem{shadow1}
K.~Akiyama et al. (Event Horizon Telescope),
Astrophys.~J.~Lett., \textbf{875}, L1 (2019).
\bibitem{shadow2}
K.~Akiyama et al. (Event Horizon Telescope),
Astrophys.~J.~Lett., \textbf{930}, L17 (2022).

\bibitem{disk1}
I.~D. Novikov, K.~S. Thorne, Black Holes (Les Astres Occlus), New York, 343 (1973).
\bibitem{disk2}
D.~N. Page, K.~S. Thorne, Astrophys. J. \textbf{191}, 499 (1974).
\bibitem{disk3}
N.~I. Shakura and R.~A. Sunyaev, Astron. Astrophys. \textbf{24}, 337 (1973).

\bibitem{ts0}
E.~Kurmanov, K.~Boshkayev, R.~Giamb{\`o}, T.~Konysbayev, O.~Luongo, D.~Malafarina and H.~Quevedo, Astrophys. J. \textbf{925}, 210 (2022).
\bibitem{ts1}
K.~Boshkayev, T.~Konysbayev, Y.~Kurmanov, O.~Luongo, M.~Muccino, H.~Quevedo and A.~Urazalina, Eur. Phys. J. Plus \textbf{139}, 273 (2024).
\bibitem{ts3}
G.~Mustafa, S.~K.~Maurya, A.~Ditta, S.~Ray and F.~Atamurotov, Eur. Phys. J. C \textbf{84}, 690 (2024).
\bibitem{ts4}
S.~Patra, B.~R.~Majhi and S.~Das, JCAP. \textbf{060}, 01 (2024).
\bibitem{ts5}
L.~A.~S\'anchez, Eur. Phys. J. C. \textbf{84}, 635 (2024).
\bibitem{ts6}
Y.~H.~Jiang and T.~Wang, Phys. Rev. D. \textbf{110}, 103009 (2024).
\bibitem{ts15}
M.~{\v{C}}emelji{\'c}, W.~Klu{\'z}niak, R.~Mishra and M.~Wielgus, Astrophys. J. \textbf{981}, 69 (2025).
\bibitem{ts16}
S.~M.~Hoseyni, J.~Ghanbari and M.~Moeen Moghaddas, Astrophys. Space Sci. \textbf{370}, 52 (2025).
\bibitem{ts21}
Y.~Ouyang, X.~Zhou, S.~Chen and J.~Jing, JCAP \textbf{08}, 094 (2025).
\bibitem{ts23}
S.~Faraji, Eur. Phys. J. C \textbf{85}, 148 (2025).

\bibitem{t1s7}
H.~B.~Zheng, M.~Q.~Wu, G.~P.~Li and Q.~Q.~Jiang, Eur. Phys. J. C \textbf{85}, 46 (2025).
\bibitem{t1s8}
Y.~Hou, Z.~Zhang, H.~Yan, M.~Guo and B.~Chen, Phys. Rev. D. \textbf{106}, 064058 (2022).
\bibitem{t1s9}
J.~Peng, M.~Guo and X.~H.~Feng, Chin. Phys. C. \textbf{45}, 085103 (2021). 
\bibitem{t1s10}
K.~J.~He, G.~P.~Li, C.~Y.~Yang and X.~X.~Zeng, Eur. Phys. J. C \textbf{85}, 662 (2025).
\bibitem{t1s11}
Z.~Li, Eur. Phys. J. C \textbf{85}, 514 (2025).
\bibitem{t1s12}
D.~Zhang, G.~Fu, X.~J.~Wang, Q.~Pan, X.~M.~Kuang and J.~P.~Wu, Eur. Phys. J. C \textbf{85}, 1051 (2025).
\bibitem{t1s13}
Z.~Li and X.~K.~Guo, Eur. Phys. J. C \textbf{85}, 679 (2025).
\bibitem{t1s14}
Y.~Meng, X.~J.~Wang, Y.~Z.~Li and X.~M.~Kuang, Eur. Phys. J. C \textbf{85}, 627 (2025).
\bibitem{t1s17}
S.~Q.~Liu and J.~H.~Huang, Phys. Rev. D \textbf{112}, 064090 (2025).
\bibitem{t1s18}
M.~Fathi, Phys. Dark Univ. \textbf{50}, 102069 (2025).
\bibitem{t1s19}
J.~Chen and J.~Yang, Eur. Phys. J. C \textbf{85}, 512 (2025).
\bibitem{t1s20}
Y.~Wu, Z.~Cai, Z.~Ban, H.~Feng and W.~Q.~Chen, Eur. Phys. J. C \textbf{85}, 1085 (2025).
\bibitem{t1s22}
K.~Kobialko, D.~Gal'tsov and A.~Molchanov, Phys. Rev. D \textbf{112}, 044039 (2025).
\bibitem{t1s24}
G.~J.~Olmo, J.~L.~Rosa, D.~Rubiera-Garcia, A.~Rueda and D.~S{\'a}ez-Chill{\'o}n G{\'o}mez, Phys. Rev. D \textbf{112}, 084059 (2025).
\bibitem{t1s25}
G.~N.~Gyulchev, D.~D.~Doneva, V.~O.~Deliyski, P.~G.~Nedkova and S.~S.~Yazadjiev, Phys. Rev. D \textbf{114}, 024074 (2026).
\bibitem{t1s26}
Z.~Y.~Zhang, X.~Q.~Li, H.~P.~Yan and X.~J.~Yue, Eur. Phys. J. C \textbf{86}, 945 (2026).
\bibitem{bz}
R.~D.~Blandford and R.~L.~Znajek, Mon. Not. R. Astron. Soc. \textbf{179}, 433 (1977).
\bibitem{MR}
L. Comisso and F. A. Asenjo, Phys. Rev. D \textbf{103}, 023014 (2021).
\bibitem{MR0}
S. Koide and K. Arai, Aprophys. J. \textbf{682}, 1124 (2008).
\bibitem{mr1}
 W. Daughton, V. Roytershteyn, B. Albright, H. Karimabadi, L. Yin, and K. J. Bowers, Phys. Rev. Lett. \textbf{103}, 065004 (2009).
\bibitem{mr2}
A. Bhattacharjee, Y.-M. Huang, H. Yang, and B. Rogers, Phys. Plasmas \textbf{16}, 112102 (2009).
\bibitem{mr3}
K. Parfrey, A. Philippov and B. Cerutti, Phys. Rev. Lett. \textbf{122}, 035101 (2019).
\bibitem{mr4}
S. S. Komissarov, Mon. Not. Roy. Astron. Soc. \textbf{359}, 801 (2005).
\bibitem{mr5}
B. Ripperda, F. Bacchini and A. Philippov, Astrophys. J. \textbf{900}, 100 (2020).
\bibitem{mr6}
A. Bransgrove, B. Ripperda and A. Philippov, Phys. Rev. Lett. \textbf{127}, 055101 (2021).
\bibitem{mr7}
L. Comisso, M. Lingam, Y.-M. Huang, and A. Bhattacharjee,  Phys. Plasmas \textbf{23}, 100702 (2016)
\bibitem{mr8}
D. A. Uzdensky and N. F. Loureiro, Phys. Rev. Lett. \textbf{116}, 105003 (2016)
\bibitem{mr9}
L. Comisso, M. Lingam, Y.-M. Huang, and A. Bhattacharjee, Astrophys. J. \textbf{850}, 142 (2017).
\bibitem{mrr1}
S. W. Wei, H. M. Wang, Y. P. Zhang, and Y. X. Liu, JCAP \textbf{04}, 050 (2022).
\bibitem{mrr2}
W. Liu, Astrophys. J. \textbf{925}, 149 (2022).
\bibitem{mrr3}
M. Khodadi, Phys. Rev. D \textbf{105}, 023025 (2022).
\bibitem{mrr4}
A. Carleo, G. Lambiase, and L. Mastrototaro, Eur. Phys. J. C \textbf{82}, 776 (2022).
\bibitem{mrr5}
C.~H.~Wang, C.~Q.~Pang and S.~W.~Wei, Phys. Rev. D \textbf{106}, 124050 (2022).
\bibitem{mrr6}
Z.~Li, X.~K.~Guo and F.~Yuan, Phys. Rev. D \textbf{108}, 044067 (2023).
\bibitem{mrr7}
Z.~Li and F.~Yuan, Phys. Rev. D \textbf{108}, 024039 (2023).
\bibitem{mrr8}
X.~Ye, C.~H.~Wang and S.~W.~Wei, JCAP \textbf{12}, 030 (2023).
\bibitem{mrr9}
M.~Khodadi, D.~F.~Mota and A.~Sheykhi, JCAP \textbf{10}, 034 (2023).
\bibitem{mrr10}
S.~Shaymatov, M.~Alloqulov, B.~Ahmedov and A.~Wang, Phys. Rev. D \textbf{110}, 044005 (2024).
\bibitem{mrr11}
S.~Shaymatov, Phys. Rev. D \textbf{110}, 044042 (2024).
\bibitem{mrr12}
S.~J.~Zhang, Phys. Rev. D \textbf{109}, 084066 (2024).
\bibitem{mrr13}
B.~Chen, Y.~Hou, J.~Li and Y.~Shen, Phys. Rev. D \textbf{110}, 063003 (2024).
\bibitem{mrr14}
S.~J.~Zhang, JCAP \textbf{07}, 042 (2024).
\bibitem{mrr15}
S.~Rodriguez, A.~Sidler, L.~Rodriguez and L.~R.~Ram-Mohan, Phys. Dark Univ. \textbf{48}, 101961 (2025).
\bibitem{mrr16}
T.~Xamidov, S.~Shaymatov, P.~Sheoran and B.~Ahmedov, Eur. Phys. J. C \textbf{84}, 1300 (2024).
\bibitem{mrr17}
Z.~Y.~Fan, Y.~Li, F.~Zhou and M.~Guo, Phys. Rev. D \textbf{110}, 104044 (2024).
\bibitem{mrr18}
Y.~Shen, H.~Y.~YuChih and B.~Chen, Phys. Rev. D \textbf{110}, 123010 (2024).
\bibitem{mrr19}
F.~Long, S.~Wang, S.~Chen and J.~Jing, Eur. Phys. J. C \textbf{85}, 26 (2025).
\bibitem{mrr20}
J.~S.~Santos, V.~Cardoso and J.~Nat{\'a}rio, Phys. Rev. D \textbf{110}, 124054 (2024).
\bibitem{mrr21}
Y.~Shen and H.~Y.~YuChih, Phys. Rev. D \textbf{111}, 023003 (2025).
\bibitem{mrr22}
X.~X.~Zeng and K.~Wang, Phys. Rev. D \textbf{112}, 064080 (2025).
\bibitem{mrr23}
Z.~Cheng, S.~Chen and J.~Jing, Eur. Phys. J. C \textbf{85}, 1130 (2025).
\bibitem{mrr24}
X.~X.~Zeng and K.~Wang, Phys. Rev. D \textbf{112}, 064032 (2025).
\bibitem{mrr25}
J.~T.~Yao, K.~J.~He, Z.~C.~Lin and H.~Yu, Phys. Lett. B \textbf{878}, 140562 (2026).
\bibitem{mrr26}
I.~Eshtursunov and S.~Shaymatov, Eur. Phys. J. C \textbf{86}, 981 (2026).
\bibitem{mrr27}
J.~Y.~Liu, B.~Zhao and C.~H.~Wang, Phys. Lett. B \textbf{879}, 140571 (2026).
\bibitem{ns1}
E.~Agol and J.~Krolik, Astrophys. J. \textbf{528}, 161 (2000).
\bibitem{ns2}
J.~R.~Weaver and K.~Horne, Mon. Not. Roy. Astron. Soc. \textbf{512}, 899 (2022).
\bibitem{bard}
J. M. Bardeen, W. H. Press, and S. A. Teukolsky, Astrophys.
J. 178, 347 (1972).
\bibitem{r1}
M.~Flock, N.~Dzyurkevich, H.~Klahr, N.~J.~Turner and T.~Henning, Astrophys. J. \textbf{735}, 122 (2011).






\end{thebibliography}
\end{document}